\documentclass[fleqn,usenatbib]{mnras}

\usepackage[T1]{fontenc}
\usepackage{ae,aecompl}

\usepackage{palatino,url,graphicx,wrapfig,array,setspace,amsmath,amssymb,fancyhdr,multirow,lscape,appendix,rotating}
\usepackage{graphics,epsfig,txfonts}

\usepackage{longtable}
\usepackage{amssymb}
\usepackage{xcolor}
\usepackage{color}
\usepackage{lipsum}
\usepackage{textcomp}
\usepackage{threeparttable}
\usepackage{enumerate}
\usepackage{placeins}
\usepackage{float}
\usepackage[normalem]{ulem}
\usepackage{soul}
\usepackage{blindtext}

\newcommand{\nicer}{\textit{NICER}}
\newcommand{\chandra}{{\it Chandra}}

\newcommand{\xmm}{{\it XMM-Newton}}

\newcommand{\swift}{{\it Swift}}

\newcommand{\nustar}{{\it NuSTAR}}
\newcommand{\nus}{{\it NuSTAR}}

\newcommand{\ergs}[1]{$\times 10^{#1}$ erg s$^{-1}$}

\newcommand{\ngc}{NGC\,300\,ULX1\xspace}

\newcommand{\eqb}{\begin{eqnarray}}
\newcommand{\eqe}{\end{eqnarray}}

\title[NGC 300 ULX1: spin evolution]{NGC 300 ULX1: spin evolution, super-Eddington accretion and outflows}

\author[G.~Vasilopoulos et al.]{
G.~Vasilopoulos$^1$\thanks{E-mail: georgios.vasilopoulos@yale.edu},
M. Petropoulou$^2$,
F. Koliopanos$^{3}$,
P.~S.~Ray$^{4}$,
C.~B.~Bailyn$^{1}$,
 \newauthor
F.~Haberl$^{5}$,
K.~Gendreau$^{6}$
\\
$^1$Yale University, PO Box 208101, New Haven, CT 06520-8101, USA \\
$^2$Department of Astrophysical Sciences, 
Princeton University, 4 Ivy Lane, Princeton, NJ 08544, USA \\
$^3$IRAP, CNRS, 9 avenue du Colonel Roche, BP 44346, F-31028 Toulouse Cedex 4, France\\
$^4$Space Science Division, U.S. Naval Research Laboratory, Washington, DC 20375, USA\\
$^5$Max-Planck-Institut für Extraterrestrische Physik, Giessenbachstra{\ss}e, 85748 Garching, Germany\\
$^6$X-Ray Astrophysics Laboratory, NASA Goddard Space Flight Center, Greenbelt, MD 20771, USA
}

\date{Accepted XXX. Received YYY; in original form ZZZ}

\pubyear{2015}

\begin{document}
\label{firstpage}
\pagerange{\pageref{firstpage}--\pageref{lastpage}}
\maketitle

\begin{abstract}
\ngc is an ultra-luminous X-ray pulsar, showing an unprecedented spin evolution, from about 126~s to less than 20~s in only 4 years, consistent with steady mass accretion rate. Following its discovery we have been monitoring the system with \swift\ and \nicer\ to further study its properties.
We found that even though the observed flux of the system dropped by a factor of $\gtrsim$20, the spin-up rate remained almost constant. A possible explanation is that the decrease in the observed flux is a result of increased absorption of obscuring material due to outflows or a precessing accretion disc.
\end{abstract}




\section{Introduction}
Ultra-luminous X-ray sources (ULXs) are binary systems that emit radiation in excess of the Eddington limit as computed for accretion onto a stellar-mass compact object  \citep{2017ARA&A..55..303K}.
Although super-Eddington mass transfer rates have been postulated for ULX systems \citep[hosting black-holes; e.g.][]{2001ApJ...552L.109K}, the first undisputed evidence was provided by the discovery of the first ultra-luminous X-ray pulsar (ULXP) M82 X-2 \citep{2014Natur.514..202B}. 
In addition, it has been shown that the X-ray spectral properties of pulsating and non-pulsating ULXs are consistent with a neutron star (NS) being the central engine \citep{2017A&A...608A..47K} and  accreting at super-Eddington rates.

Given the super-Eddington mass accretion rates in ULXPs, it is expected that a fraction of the total accreted mass will be lost through outflows, launched from the inner accretion disc  \citep{1973A&A....24..337S}. These outflows can be optically thick to the  X-ray radiation produced by the NS, thus creating an obscuring envelope around it \citep{2007MNRAS.377.1187P}. However, the outflows are not spherical and X-ray  radiation can still escape from a central hollow cone region 
\citep{2017MNRAS.468L..59K,2019MNRAS.485.3588K}.

For three of the known ULXPs, orbital periods have been derived from pulsar timing, while super-orbital periodicities have been evident in their X-ray light curves \citep[e.g.][]{2006ApJ...646..174K,2014Natur.514..198M}. 
M82 X-2 has an orbital period of $\sim$2.5 d and a $\sim$64 d super-orbital period \citep{2014Natur.514..202B,2019ApJ...873..115B}. NGC\,5907\,ULX1 has a 5.3 d and 78 d orbital and super-orbital period respectively \citep{2017Sci...355..817I,2017ApJ...834...77F}.
NGC 7793 P13 has an orbital period of $\sim$64 d, while its X-ray light-curve has a $\sim$67 d super-orbital periodicity \citep[e.g.][]{2018A&A...616A.186F}.  
The observed X-ray flux ($F_{\rm X}$) during these super-orbital phases can vary by a factor of 100, with no evidence of spectral changes expected from   accretor-to-propeller transitions \citep{1975A&A....39..185I,1996ApJ...457L..31C,2018A&A...610A..46C}. A possible scenario to explain the flux changes on super-orbital timescales is the obscuration from a precessing accretion disc  \citep{2017ApJ...834...77F,2018MNRAS.475..154M}.
However, the latter has never been confirmed by any timing study of their pulse period evolution.

\ngc is a recently discovered ULXP \citep{2018MNRAS.476L..45C}, located at a distance of 1.88 Mpc \citep{2005ApJ...628..695G}. 
The system was classified as a super-nova impostor, after it became active in 2010 \citep[SN 2010da;][]{2011ApJ...739L..51B}.
Its ULXP classification was based on the analysis of data obtained by simultaneous \xmm\ and \nus\ observations in December 2016, which yielded a spin period ($P$) of $\sim$31~s and an unabsorbed X-ray luminosity (in the 0.3--30 keV band) $L_{\rm X}\sim$4.7~\ergs{39} \citep{2018MNRAS.476L..45C}. 
Analysis of the X-ray spectra of \ngc has shown that its intrinsic X-ray luminosity  between 2010 and early 2018 remained constant (within a factor of $\sim$3), with any large changes in its observed flux being attributed to a variable absorption component \citep{2018MNRAS.476L..45C,2018A&A...620L..12V,2019A&A...621A.118K}. Analysis of archival X-ray data revealed a remarkable spin up of the NS, whose spin period changed from $\sim$126~s to $\sim$19~s between November 2014 and April 2018 \citep{2018A&A...620L..12V}. Further X-ray monitoring of \ngc during the first half of 2018 with \nicer~revealed only small changes in the observed $F_{\rm X}$ and confirmed a constant spin-up rate  \citep{2018arXiv181109218R}.

In this work, we study the evolution of \ngc  using X-ray monitoring observations that cover the time period between August 2018 and December 2018. We show that the observed X-ray flux of \ngc dropped by a factor of $\gtrsim$20-30 from its peak value in 2018, but the NS spin-up rate remained roughly constant (\S\ref{sec:data}). 
The latter requires a constant accretion rate, and thus a steady energy release (i.e. intrinsic $L_{\rm X}$) assuming the radiative efficiency remains the same.
Our results can be understood in the context of super-Eddington accretion onto a magnetized NS (\S\ref{sec:model}), with the required absorption being caused by either obscuration from a precessing accretion disc  or mass outflows launched from the inner parts of the disc  due to very high  accretion rates (\S\ref{sec:discuss}).

\section{X-ray monitoring observations}\label{sec:data}

Our findings are based on \swift/XRT and \nicer~observations of \ngc that were obtained mainly in 2018.
Between January 2018 and July 2018 the X-ray flux of \ngc gradually declined by a factor of $\sim$2, but still remained at a super-Eddington level \citep{2018arXiv181109218R}. In August-September 2018 the observed $F_{\rm X}$ rapidly declined and the system was no longer visible with either \nicer\ or \swift/XRT. The latest spin period measurement, on August 21 2018, yielded  $P=17.52\pm0.04$~s.

In November 2018, \swift/XRT monitoring observations (PI: Kennea, J) indicated a re-brightening of the system. We thus requested \nicer\ target of opportunity (ToO) observations (PI: Vasilopoulos, G) to measure the NS's pulse period. \nicer\ measured a spin period of $16.58\pm0.04$~s on November 28 2019 (MJD 58451), revealing a spin change of $\sim1$~s within $\sim 100$ days.  

\begin{figure*}
  \resizebox{\hsize}{!}{
    \includegraphics[angle=0,clip=]{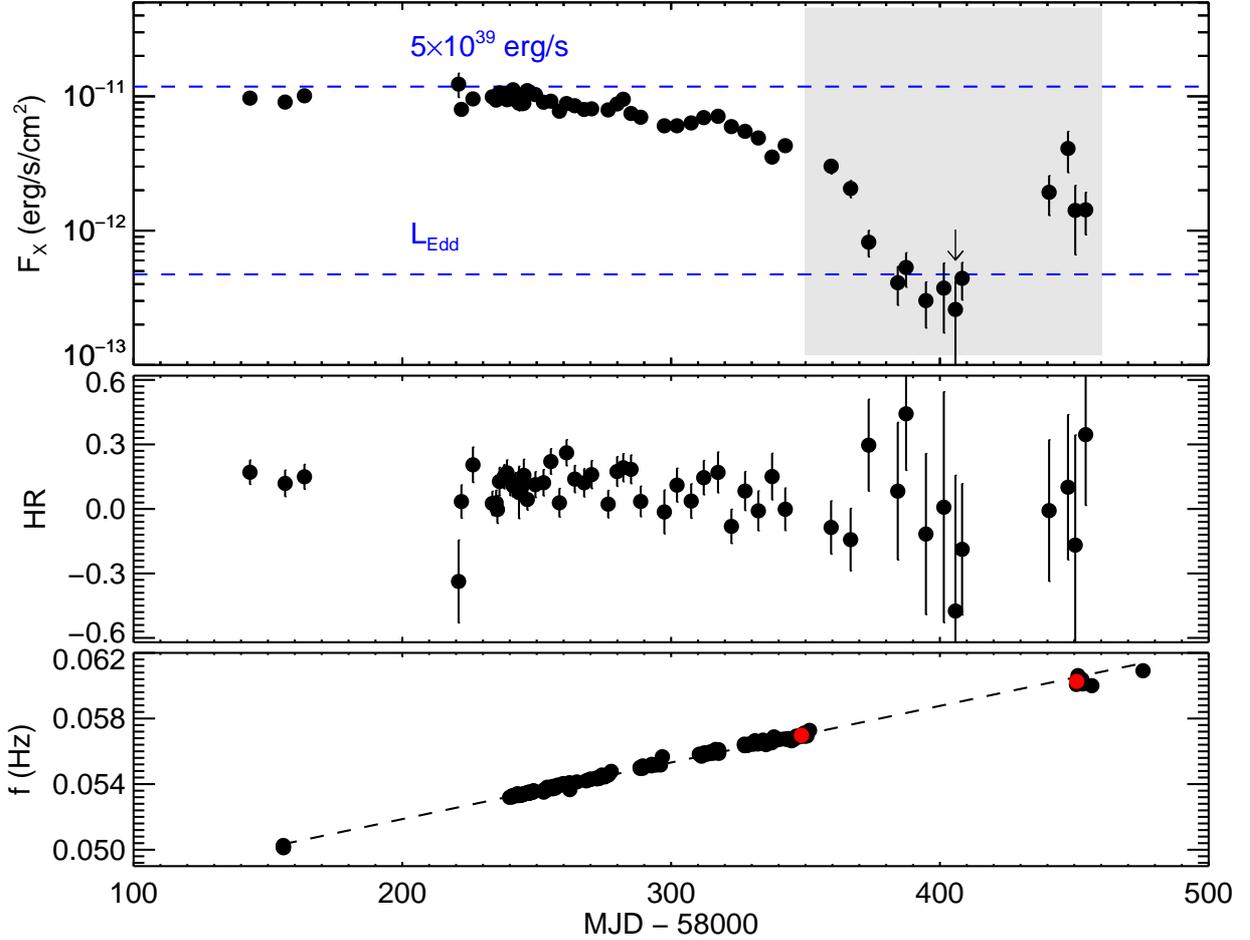}}
        \vspace{-0.9cm}
  \caption{\emph{Top panel:} X-ray light-curve (0.3--30 keV band) of \ngc as derived from \swift/XRT observations (black points) performed within 2018.
  We used the spectral properties derived by \citet{2018MNRAS.476L..45C}, assuming a constant absorption, to convert \swift/XRT count rates to $F_{\rm X}$ (see text).
  Horizontal dashed blue lines mark characteristic luminosity levels converted to expected flux for a distance of 1.88 Mpc. 
  The shaded area indicates the period of the rapid X-ray flux decline and recovery discussed in text. 
  \emph{Middle panel:} temporal evolution of the hardness ratio (HR) as derived from the soft (0.3--1.5 keV) and hard (1.5--10.0 keV) energy bands.  
  \emph{Bottom panel:} temporal evolution of the measured spin frequencies derived from \nicer \ observations \citep[see][]{2018arXiv181109218R}. The black dashed line denotes the best linear fit to the data, which yields $\dot{M}= 2.3\times10^{19}$ g~s$^{-1}$ (see text for details; \S \ref{sec2}). 
   Red points are the \nicer\ detections with the highest significance \citep[see ][for details]{2018arXiv181109218R} just before the decay and after the rise of the $F_{\rm X}$.
  }
  \label{fig:lx}
\end{figure*}

\subsection{Data analysis}

We used \swift/XRT to measure the X-ray flux of \ngc. \swift/XRT data products are available though the UK \swift \ science data centre\footnote{\url{http://www.swift.ac.uk/user_objects/}} \citep{2007A&A...469..379E,2009MNRAS.397.1177E}.
Basic information about spectral changes of the observed spectrum can be derived from the hardness ratio (HR). This is defined as the ratio of the difference over the sum of the number of counts in two subsequent energy bands: $\rm{HR}=(\rm{R}_{\rm{i+1}}-\rm{R}_{\rm{i}})/(\rm{R}_{\rm{i+1}}+\rm{R}_{\rm{i}})$,   
where $\rm{R}_{\rm{i}}$ is the background-subtracted count rate in a specific energy band. 

We used data obtained by \nicer\ from 2018 February 6 through 2019 February 4. to obtain accurate measurements of the spin frequency evolution of \ngc.  We followed the maximum likelihood procedure described by \citet{2018arXiv181109218R} on segments of data with gaps less than 1000 s and spanning no more than 3000 s. The points plotted in Figure 1 are the detections where the data span at least 360 seconds with a significance of at least 4.5 $\sigma$.

\subsection{Results}\label{sec:results}

\citet{2018MNRAS.476L..45C} found that the multi-epoch broadband spectra of \ngc\ are best fitted by a multi-component continuum, where the spectrum can be phenomenologically described by a power-law ($\Gamma\sim1.5$) with cut-off ($E_{\rm cut}\sim$6.6 keV, $E_{\rm fold}\sim$4.8 keV), with a soft excess (i.e., below 1.0 keV) attributed to a black body component ($kT\sim$0.17 keV).
Most importantly, the authors showed that the spectral continuum is partially absorbed by material with high column density, showing no clear low-energy cutoff but a rather complicated spectral shape \citep[see also][]{2019A&A...621A.118K}.
The \swift/XRT data do not provide enough statistics for a detailed spectral fit and, as a result, we cannot constrain  the partial absorption component.
Thus, for the conversion of \swift/XRT count rates to $F_{\rm X}$ (0.3-30.0 keV) we used the spectral properties derived by \citet{2018MNRAS.476L..45C} for the continuum spectrum, a constant absorption (accounting only for Galactic absorption), and assumed  distance of 1.88 Mpc. 
In other words, the derived X-ray light curve, shown in Fig.~\ref{fig:lx}, is not corrected for the intrinsic absorption of the system\footnote{In this context the observed flux $F_{\rm X}$, plotted in Fig.~\ref{fig:lx}, when corrected for the distance modulus of NGC 300, is the $L_{\rm X}$.}. 
Within the period August-November 2018 (see grey shaded area in Fig.~\ref{fig:lx}; MJD $\sim$58350-58450), the observed $F_{\rm X}$ of \ngc rapidly decreased. 
As already noted, due to the limited statistics we can put no constraints on the spectral change that accompanied this transition. 

Prior to this work, \citet{2018A&A...620L..12V} have investigated the spin evolution of \ngc\ by analysing archival X-ray data and searching for periodic signals using the accelerated epoch-folding method \citep{1983ApJ...266..160L}.
Both $P$ and $\dot{P}$ were measured, with small uncertainties, for 3 epochs where \xmm, \chandra, \ and \nustar \ data with high statistics were available.
The derived values of $\dot{\nu}$ are  $5.5\times10^{-10}$, $4.5\times10^{-10}$, and $3.8\times10^{-10}$ s$^{-2}$ \citep{2018A&A...620L..12V}.
We used \nicer\ monitoring data to extend these measurements.
During 2018 the evolution of the NS spin frequency follows an almost linear trend with time (see bottom panel in Fig.~\ref{fig:lx}).
To derive precise period measurements, we employed  a  maximum  likelihood  technique to  measure  pulsed  frequencies  and their significance \citep[see details][]{2011ApJS..194...17R,2018arXiv181109218R}. For this purpose, we only used ``good'' detections\footnote{These have likelihood test statistic less than 21, span of observation greater than 360 s, and more than 70\% of good exposure time within that span \citep[see also][]{2018arXiv181109218R}.}.
We fitted the time series of the NS frequencies with a Bayesian approach to linear regression \citep{2007ApJ...665.1489K} and derived a slope of $(4.031\pm0.026)\times10^{-10}$~s$^{-2}$  (90\% confidence level). 
When fitting the data with a polynomial model instead, we derive an ephemeris with ${\nu}=0.053347\pm0.000003$~s$^{-1}$,
$\dot{\nu}=(4.23\pm0.03)\times10^{-10}$~s$^{-2}$ and $\ddot{\nu}=(4.4\pm0.5)\times10^{-18}$~s$^{-3}$ at epoch MJD 58243.515.
The secular frequency evolution as derived above is in agreement ($\sim20\%$ deviation) with the instantaneous $\dot{v}$ measured by \citet{2018MNRAS.476L..45C,2018A&A...620L..12V}.

In the following section, we show that the spin evolution of the NS in \ngc during 2018 requires a roughly constant accretion rate in excess of the Eddington limit ($\sim2\times10^{18}$ g s$^{-1}$), whereas the light curve clearly shows a decrease of the observed $F_{\rm X}$. 
This controversy can be naturally resolved, if one attributes the decay of the observed $F_{\rm X}$ to an increased absorption due to extra material present in the line of sight. The absorbing material can be the result of outflows launched from the accretion disc  and/or of the disc  precession.

\section{Accretion and torques in ULX pulsars}
\label{sec:model}

The Eddington luminosity of an accreting object is obtained by equating the outward radiation pressure with the gravitational force:
\begin{equation}
L_{\rm Edd}=\frac{4{\pi}GM_{\rm NS}c}{\kappa}\approx1.5\times10^{38} m_1 \, {\rm erg \ s}^{-1},
\label{eq3}
\end{equation}
where $G$ is the gravitational constant, $\kappa=0.2(1+X)~{\rm cm^2 g^{-1}}$ is the Thomson opacity, $X$ is the hydrogen mass fraction for solar abundances ($X=0.7$), and $m_1\equiv M_{\rm  NS}/M_{\odot}$ is the NS mass (in units of the solar mass). 

At low mass accretion rates, the disc is locally sub-Eddington and its inner part is gas-pressure dominated (regime 1). For mass accretion rates exceeding the critical rate $\dot{M}_{\rm Edd}$:
\begin{equation}
\dot{M}_{\rm Edd} = \frac{48 \pi G M_{\rm NS}}{c\kappa}\simeq2\times 10^{18} m_1\, {\rm g \ s}^{-1},
\label{eq:Medd}
\end{equation}
part of the dissipated energy is used to launch mass outflows from the inner part of the accretion disc, thus leading to a reduced accretion rate onto the compact object.  
This occurs inside the spherization radius $R_{\rm sph}$, where the disc thickness becomes comparable to its radius \citep{1973A&A....24..337S} or, equivalently, the disc  luminosity becomes equal to $L_{\rm Edd}$ \citep[see eq.~18 in][]{2007MNRAS.377.1187P}:
\begin{equation}
R_{\rm sph} \approx 10 \frac{GM_{\rm NS}\dot{m}_0}{c^2} \simeq 15 \, m_1\dot{m}_0 \, {\rm km},
\label{eq4}
\end{equation}
where  
$\dot{m}_0$ is the mass accretion rate at $R_{\rm sph}$ in units of $\dot{M}_{\rm Edd}$.
In this regime, the outflows launched from within the spherization radius are re-configuring accretion in a way that the local disc  accretion rate is sub-Eddington \citep{1973A&A....24..337S}. In the classical mass-loss model of \citet{1973A&A....24..337S}, the mass accretion rate at $R<R_{\rm sph}$ can be  written as:
\begin{equation}
\dot{M}(R)\simeq\frac{R}{R_{\rm sph}}\dot{m}_0\dot{M}_{\rm Edd}. 
\label{eq5}
\end{equation}
This is only an approximate relation, as it does not take into account the effects of heat advection in the disc. Inclusion of the latter results in a more gradual decrease of $\dot{M}$ with radius than the one dictated by eq.~(\ref{eq5}) \citep[][]{2007MNRAS.377.1187P, Anna2019,2019MNRAS.484..687M}. 

In contrast to non-magnetized accreting objects, the accretion discs around magnetized NSs do not extend to the innermost stable orbit, but they are truncated at much larger radii due to the interaction with the NS magnetic field. The magnetospheric radius provides an estimate of the disc  inner radius \citep{1977ApJ...217..578G}:
\begin{equation}
R_{\rm M} = \xi \left(\frac{R_{\rm NS}^{12}B^4}{2GM_{\rm NS}\dot{M}^2}\right)^{1/7},
\label{eq1}
\end{equation}
where $R_{\rm NS}$ is the neutron star radius and $\xi\sim 0.5$ \citep{2018A&A...610A..46C}.
For typical $B$ values in X-ray pulsars (e.g., $10^{12}$ G), very high mass accretion rates are required 
(e.g., $\dot{m}_0>10$) to make $R_{\rm sph}>R_{\rm M}$ (regime 2).

Although heat advection is operating in the disc even at sub-Eddington rates, where no outflows are present, it begins to play an increasingly important role in the energy balance of the inner disc at super-Eddington rates, i.e., $\dot{m}_0\gtrsim20$ (regime 3) and eventually becomes the dominant process at extremely high rates, i.e., $\dot{m}_0 \gtrsim 100$ for typical X-ray pulsar magnetic fields \citep[see Fig.~12 of][]{Anna2019}\footnote{ Note that \citet{Anna2019} normalize the accretion rates to $4\pi G M_{\rm_{NS}}/c\kappa$, which introduces a factor of 12 difference with the normalization adopted here.}. Because of the heat advection, the radiation energy flux transported by diffusion in the vertical direction is less than the one released locally in the disc  \citep{2019MNRAS.484..687M}. The advection process effectively leads to a reduced mass loss from the disc. The  outflow rate in this regime is a constant fraction of the mass accretion rate at $R_{\rm sph}$ and cannot exceed $\sim 50\%-60\%$ \citep[see e.g., Fig.~3 in][]{2019MNRAS.484..687M}.

\begin{figure}
  \resizebox{\hsize}{!}{
       \includegraphics[angle=0,clip=]{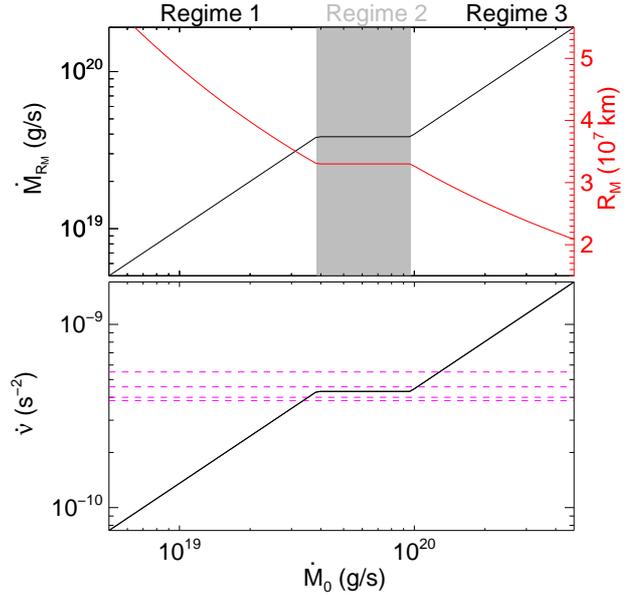}
     }
 \vspace{-0.5cm}
  \caption{\emph{Top panel:} Schematic dependence of the magnetospheric (inner disc) radius $R_{\rm M}$ (red line) and the  mass accretion rate at that radius $\dot{M}(R_{\rm M})$ (black line) as a function of the mass accretion rate at the spherization radius $\dot{M}_0$  \citep[for detailed calculations, see][]{2019MNRAS.484..687M, Anna2019}. The spherization ($R_{\rm sph}$) and magnetospheric  radii are defined by eqs.~(\ref{eq4}) and (\ref{eq:Rm}), respectively. The condition $R_{\rm sph}=R_{\rm M}$ marks the transition between regimes 1 and 2. For the definitions of the accretion regimes indicated in the plot, see  \S\ref{sec:model}. \emph{Bottom panel:} Derived spin-up rate using eq.~(\ref{eq2}). Horizontal lines mark the various $\dot{\nu}$ measurements based on \xmm, \chandra, \nustar, \ and \nicer\ data: $5.5\times10^{-10}$, $4.5\times10^{-10}$, $4.0\times10^{-10}$, and $3.8\times10^{-10}$ s$^{-2}$ \citep{2018A&A...620L..12V,2018arXiv181109218R}.}
  \label{fig:regimes}
\end{figure}

In what follows, we interpret our findings (see  Fig.~\ref{fig:lx} and \S~\ref{sec:results}) in the context of the three regimes described above and schematically shown in Fig.~\ref{fig:regimes}.

\subsection{Spin-evolution without outflows}
\label{sec2}

The spin evolution of a sub-Eddington accreting NS can be theoretically predicted, if two parameters are  known,  the accretion rate and the surface magnetic field of the NS \citep{1995ApJ...449L.153W}. 
In X-ray pulsars, material is deposited onto the magnetic pole of the NS, forming the so-called accretion column \citep[e.g.][]{2007ApJ...654..435B,2015MNRAS.454.2539M}.
The bolometric X-ray luminosity of the system, which originates from the accretion column, can be converted to a mass accretion rate $\dot{M}$ assuming some efficiency $\eta_{\rm eff}$ (i.e., $L_{\rm X}\approx \eta_{\rm eff}\dot{M}c^2$).
This is generally assumed to be the efficiency with which gravitational energy is converted to radiation, namely $L_{\rm X}=GM_{\rm NS}\dot{M}/R$. For $R=R_{\rm NS}=10^6$~cm and $M_{\rm NS}=1.4M_{\odot}$, one finds $L_{\rm X}\approx0.2\dot{M}c^2$. Henceforth, we adopt $\eta_{\rm eff}=0.2$.

The induced torque due to the mass accretion is 
$N_{\rm acc}\approx\dot{M}\sqrt{GM_{\rm NS}R_{\rm M}}$. 
The total torque can be expressed in the form of $N_{\rm tot}=n(\omega_\mathrm{fast})N_{\rm acc}$ where $n(\omega_\mathrm{fast})$ is a dimensionless function that accounts for the coupling of the magnetic field lines to the accretion disc  and takes the value $\approx7/6$ for slow rotators  \citep[for more details see][]{1995ApJ...449L.153W,2016ApJ...822...33P}.
The spin-up rate of the NS is then given by: 
\begin{equation}
\dot{v}=\frac{n(\omega_\mathrm{fast})}{{\rm 2 \pi} I_{\rm NS}} \dot{M} \sqrt{G M_{\rm NS} R_{\rm M}},
\label{eq2}
\end{equation}
where $I_{\rm NS} \simeq (1-1.7)\times10^{45}$~g cm$^{2}$ is the moment of inertia of the NS \citep[e.g.,][]{2015PhRvC..91a5804S}.

For an almost constant mass accretion rate (within a factor of $\sim2$), the standard torque model can explain the NS spin period evolution for the entire period prior to MJD 58300 \citep[][]{2018A&A...620L..12V}, when the X-ray flux of \ngc started to decline. 
Observations of \ngc performed by \xmm, \nustar, \ and \chandra\ have been used to determine the system's $L_{\rm X}$ and $\dot{v}$ at different epochs, thus allowing the estimation of the NS magnetic field, which was found to be $B\simeq6\times10^{12}$~G \citep[][]{2018A&A...620L..12V}. 
Upon inserting the average spin-up rate for \ngc\ as inferred by the fit to the \nicer\ data ($\dot{v}=4.031\times10^{-10}$~s$^{-2}$) into equations (\ref{eq1})-(\ref{eq2}) and setting $\xi=0.5$, $n(\omega_{\rm fast})=7/6$, $m_1=1.4$ and $I_{\rm NS,45}=1.3$, we find 
$R_{\rm M}\simeq800$~km and $\dot{M}(R_{\rm M})\simeq 2.3\times10^{19}$~g s$^{-1}$ (see first column in Table \ref{tab1}).

\begin{figure}
  \resizebox{\hsize}{!}{
  \includegraphics[angle=0,clip=]{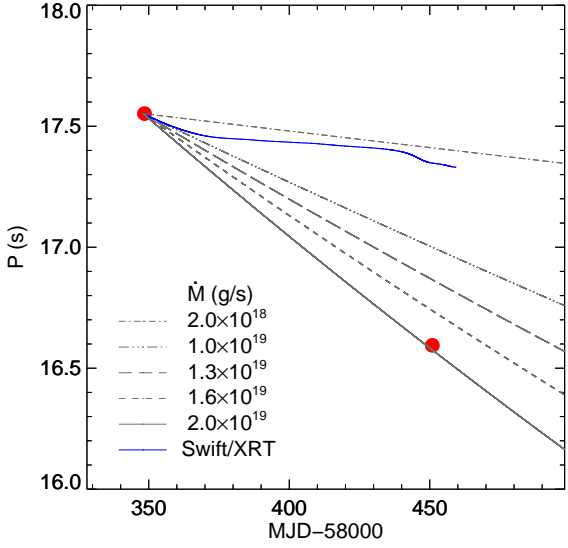}
     }
 \vspace{-0.9cm}
  \caption{Predicted evolution of NS spin period $P$ based on the \citet{1995ApJ...449L.153W} model for various accretion rates marked on the plot.
  Two red points mark the measured spin period from the \nicer \, observations, as shown also in the lower panel of Fig.~ \ref{fig:lx}. 
  Black lines denote the evolution of $P$ after the \nicer \, measurement (first red point) for different constant accretion rates (at $R_{\rm M}$) marked on the plot. The evolutionary track predicted for a variable $\dot{M}$ that follows the observed \swift/XRT count rate (shaded area in upper panel of Fig. \ref{fig:lx}) is plotted with a solid blue line. This solution is obtained using the best fit spectral model of \citet{2018MNRAS.476L..45C} with no changes in absorption and $\eta_{\rm eff}=0.2$. 
  }
  \label{fig:ps}
\end{figure}

We next explored if the same model can reproduce the spin-evolution of the NS for the 100-day period during which the observed $F_{\rm X}$ rapidly decayed and recovered (see grey shaded region in Fig.~\ref{fig:lx}). For this purpose, we solved eq.~(\ref{eq2}) assuming a constant magnetic field strength of $B=6\times10^{12}$~G and a radiative efficiency of $\eta_{\rm eff}=0.2$.
We performed our calculations for (i) various constant values of the accretion rate (or equivalently intrinsic $L_{\rm X}$) and (ii) a variable accretion rate that best describes the observed temporal evolution of $F_{\rm X}$, assuming no correction due to obscuration or extreme absorption. The derived spin-period evolutionary tracks for the two cases are plotted in Fig. \ref{fig:ps}. It is clear that the observed spin evolution does not match the one predicted for a variable mass accretion rate that matches the observed rapid decay of the X-ray flux (solid blue line).
On the contrary, 
the observed spin evolution demands a constant super-Eddington accretion rate ($\dot{M}\simeq1.6\times10^{19}$~g s$^{-1}$) that translates to an 
absorption-corrected luminosity $L_{\rm X}\sim$3\ergs{39}, in agreement with the findings of \citet{2018A&A...620L..12V}.

\subsection{Spin-evolution with outflows}
\label{sec3}

Here, we expand the calculations for the spin evolution of \ngc presented by \citet{2018A&A...620L..12V}, by taking into account the effects of outflows expected during super-Eddington accretion
\citep{1973A&A....24..337S}.  A major difference of the following analysis with that presented in \S\ref{sec2}, is that (under certain conditions) one can estimate the unknown parameters of the system ($R_{\rm M}$, $B$, $\dot{M}(R_{\rm M})$) from just one observable (i.e., the spin-up rate $\dot{\nu}$). 

Equation \ref{eq4} suggests that if $\dot{m}_0$ is sufficiently high so that $R_{\rm sph}>R_{\rm M}$, by further increasing $\dot{m}_0$, both $R_{\rm M}$ and $\dot{M}(R_{\rm M})$ remain constant and all the excess mass is lost through outflows \citep{1982SvA....26...54L}.
These arguments have provided the baseline for the recent study of \citet{2017MNRAS.468L..59K}, where it was shown that by just measuring the spin-up rate of the NS in ULXPs one can estimate $B$, $R_{\rm M}$ and $\dot{M}(R_{\rm M})$, without any knowledge of its actual X-ray luminosity (see \S\ref{sec2} for comparison). 
Using equations (\ref{eq4}), (\ref{eq5}), (\ref{eq1}), and (\ref{eq2}) we obtain $R_{\rm M}$, $B$, and $\dot{M}(R_{\rm M})$: 
\begin{equation}
R_{\rm M}\simeq 119 \, m_1^{-1/3} I_{\rm NS, 45}^{2/3} \,  \dot{\nu}_{-10}^{2/3} \left[n(\omega_{\rm fast}\right]^{-2/3} {\rm km}
 \label{eq:Rm}
\end{equation}

\begin{equation}
B\simeq 3.9\times10^{10} \, \xi^{-7/4} m_1^{-1/2} R_{\rm NS, 6}^{-3} \, I_{\rm NS, 45}^{3/2} \,  \dot{\nu}_{-10}^{3/2}\left[n(\omega_{\rm fast})\right]^{-3/2}{\rm G}
   \label{eq:B}
   \end{equation}
 
\begin{equation}
\dot{M}(R_{\rm M})\simeq 1.6\times10^{19} \,m_1^{-1/3}\, I_{\rm NS, 45}^{2/3} \, \dot{\nu}_{-10}^{2/3} \left[n(\omega_{\rm fast})\right]^{-2/3} {\rm g \, s}^{-1},
\label{eq:dotMRm}
\end{equation}
where $R_{\rm NS,6}\equiv R_{\rm NS}/10^6~{\rm cm}$, $\dot{\nu}_{-10}\equiv \dot{\nu}/10^{-10} \, {\rm s}^{-2}$ and $I_{\rm NS,45}\equiv I_{\rm NS}/10^{45} \, {\rm g \, cm}^2$. By inserting the average spin-up rate for \ngc\ as inferred by the fit to the \nicer\ data ($\dot{v}=4.031\times10^{-10}$~s$^{-2}$) into equations (\ref{eq:Rm})-(\ref{eq:dotMRm}) and setting $\xi=0.5$, $n(\omega_{\rm fast})=7/6$, $m_1=1.4$ and $I_{\rm NS,45}=1.3$, we find $B\simeq 1.4\times10^{12}$ G, $R_{\rm M}\simeq300$~km and $\dot{M}(R_{\rm M})\simeq4\times10^{19}$~g s$^{-1}$ (see second column in Table~\ref{tab1}). For these parameters, the inner disc is expected to be radiation-pressure dominated \citep[see e.g., Fig.~7 in][]{2019arXiv190505593M}.

We can also compute the luminosity that would be released as the mass reaching $R_{\rm M}$ is accreted onto the NS by adopting the 
value of $\dot{M}(R_{\rm M})$ we derived above. This yields an isotropic X-ray luminosity of $6.6$\ergs{39} (for $\eta_{\rm eff}=0.2$),  which is  a factor of 2 higher than the observed luminosity attributed to the pulsating hard component of the X-ray spectrum in \ngc \citep{2019A&A...621A.118K,2018MNRAS.476L..45C}. Unless the radiative efficiency is much lower than $10\%-20\%$, no beaming is needed to explain the spectral and temporal properties of \ngc. 

Using the derived value of $R_{\rm M}$ and the condition $R_{\rm sph}> R_{\rm M}$, we can derive a lower limit on the mass accretion rate at the spherization radius, i.e., $\dot{m}_0 \gtrsim 14$. 
We can also place a rough upper limit on $\dot{m}_0$ by noting that 
$\dot{\nu}$ does not vary much (see Fig.~\ref{fig:lx}), suggesting an approximately constant accretion rate at the magnetospheric radius. The latter condition can be realized only for a narrow range of super-Eddington accretion rates (for a schematic illustration see grey-shaded region in the top panel of Fig.~\ref{fig:regimes}). This range of accretion rates was  found to be only weakly dependent on the magnetic field strength of the NS \citep[see Figs.~12-13 in][]{Anna2019}.
Beyond this range, the  time-derivative of the NS spin frequency is expected to depend more strongly on the accretion rate, as schematically shown in the bottom panel of Fig.~\ref{fig:regimes}. This statement is true regardless of the specific dependence of $R_{\rm M}$ on $\dot{m}_0$  \citep[see e.g., Fig.~12 in][]{Anna2019}, which is determined by the interplay of mass loss through outflows and advection of heat in the inner disc and is model-dependent.
As the long-term  spin evolution of \ngc is consistent with an almost constant $\dot{\nu}$ (see Fig.~\ref{fig:lx}),  we can  place a rough upper limit on the accretion rate at the spherization radius, i.e., $\dot{m}_0 \lesssim 50$ for $B\sim 10^{12}$~G.

\begin{table}
\begin{center}
\caption{Physical parameters of \ngc inferred from its spin evolution, assuming different accretion regimes.}
\label{tab1}
\begin{threeparttable}
\begin{tabular*}{\columnwidth}{p{0.3\columnwidth}p{0.3\columnwidth}p{0.3\columnwidth}}
\hline\noalign{\smallskip}  
 Parameters & Regime 1$^{(a)}$ & Regime 2$^{(b)}$ \\
\hline\noalign{\smallskip}    
$B$ [G] & 6$\times10^{12}$ fixed$^{(c)}$ &  1.4$\times10^{12}$\\
$\dot{M}(R_{\rm M})$ [g s$^{-1}$] $^{(d)}$& 2.3$\times10^{19}$ &   4.0$\times10^{19}$\\
$R_{\rm M}$ [km]&  800 & 300\\
\noalign{\smallskip}\hline\noalign{\smallskip}
\hline 
\end{tabular*}
 \tnote{(a)} $R_{\rm M}>R_{\rm sph}$; accretion without outflows.
 \tnote{(b)} The mass-accretion rate at the magnetosphere is assumed constant.
 \tnote{(c)} Fixed at the value derived by \citet{2018A&A...620L..12V}.
 \tnote{(d)} Value of an effective constant $\dot{M}(R_{\rm M})$ that would result in the observed average spin-up rate measured by the \nicer\ data. 
\end{threeparttable}
\end{center}
\end{table} 

\section{On the origin of the obscuring material}\label{sec:discuss}

We have demonstrated that the NS continues to spin up with an almost constant $\dot{\nu}$ during the period of decreasing flux (MJD 58350-58450), suggesting an almost constant accretion rate at the NS magnetospheric radius. Thus, the decrease of the observed X-ray flux cannot be intrinsic to the source but it can rather be  caused by absorption. 

Radiatively driven outflows launched from the disc  can be optically thick to the hard radiation produced by the NS, acting effectively as obscuring envelopes \citep[e.g.,][]{2007MNRAS.377.1187P,2009PASJ...61..213A}. However, radiation can still escape from a central conical region (with opening angle $\theta_{\rm c}$) that is devoid of obscuring  material \citep{2007MNRAS.377.1187P}. For a radiation-pressure dominated accretion disc, truncated at $R_{\rm M}<R_{\rm sph}$, the mass outflow rate up to a radius $R$, lying within the spherization radius, can be estimated by:
\begin{equation}
\dot{M}_{\rm out}(R) = \int_{R_{\rm M}}^R {\rm d}R' \frac{{\rm d}\dot{M}(R')}{{\rm dR'}} \approx \dot{m}_0  \dot{M}_{\rm Edd}\frac{R-R_{\rm M}}{R_{\rm sph}},
\label{eq:Mout}
\end{equation}
where eq.~(\ref{eq5}) was used. Following \citet{2007MNRAS.377.1187P} (see eqs.~(27), (28), and (30) therein), we calculate the maximum Thomson optical depths in the directions perpendicular ($\tau_{\perp,\max}$) and parallel ($\tau_{//,\max}$) to the accretion disc  plane: 
\begin{eqnarray}
\label{eq:tauperp}
\tau_{\perp,\max}&=& \frac{\tau_0 \dot{m}_0}{\beta r_{\rm sph}}\left(\sqrt{r_{\rm sph}}-\frac{r_{\rm M}}{\sqrt{r_{\rm sph}}} \right) \\
\tau_{//,\max}&=& \frac{\tau_0 \dot{m}_0}{\beta \cot\theta_{\rm c} r_{\rm sph}}\left[2\left(\sqrt{r_{\rm sph}}-\sqrt{r_{\rm M}}\right)+2r_{\rm M}\left(r_{\rm sph}^{-1/2}-r_{\rm M}^{-1/2}\right)\right. \nonumber \\
&+&\left. \sqrt{r_{\rm sph}}\left(1-\frac{r_{\rm M}}{r_{\rm sph}} \right)\right],
\label{eq:taupar}
\end{eqnarray}
where $\beta\sim 1$ is the ratio of the outflow speed to the Keplerian velocity at $R_{\rm sph}$,  $\theta_{\rm c}\sim \pi/4$, $\tau_0\equiv\sqrt{6}\kappa \dot{M}_{\rm Edd}/4\pi c R_0$, $R_0=6 G M_{\rm NS}/c^2$, and $r\equiv R/R_0$. For $R_{\rm M}=300$~km (see Table~\ref{tab1}) and $\dot{m}_0 \gtrsim 20$, we find $\tau_{//,\max}\approx \tau_{\perp,\max} \gtrsim 5$. This corresponds to a maximum column density $N_{\rm H}\gtrsim7.5\times10^{24}$~cm$^{-2}$ for a line-of-sight cutting through the outflow, and it is sufficient to explain the X-ray flux decay due to absorption \citep[see also][]{2018MNRAS.476L..45C}.

The accretion disc  around a black hole or a NS can precess due to the Lense-Thirring effect \citep{1975ApJ...195L..65B,1986ApJ...300L..63T}.
Recently, it has been postulated that this mechanism can explain the super-orbital modulation in the observed flux of ULXPs \citep{2018MNRAS.475..154M}.
According to this scenario, the drop in the observed flux is due to changes in the geometrical configuration of the inner accretion disc  and the outflow. Thus, in ULXPs the observer can only see the NS when the wind-free region of the outflow aims directly at him/her.
The timescale of the outflow precession is \citep[see eq. 5 of][]{2018MNRAS.475..154M}:
\begin{equation}
    P_{\rm prec}\approx P_{\rm NS}\frac{R^3_{\rm sph}c^2}{6 GI_{\rm NS}}\frac{1-(R_{\rm M}/R_{\rm sph})^3}{\ln{(R_{\rm sph}/R_{\rm M})}}\left(\frac{R_{\rm out}}{R_{\rm sph}}\right)^2,
\label{eq.ltp}    
\end{equation}
where $R_{\rm out}$ is the radius at which the precession of the outflow ceases. An upper limit on the precession period can be derived if $R_{\rm out}\sim R_{\rm ph}$, where the photospheric radius $R_{\rm ph}$ is estimated by the condition
 $\tau_{\perp}=1$:
\begin{equation}
R_{\rm ph}\approx \frac{\tau_0 \dot{m}_0 R_0}{\beta \sqrt{r_{\rm sph}}}\left(r_{\rm sph}-r_{\rm M}\right).
\label{eq.rph}    
\end{equation}
For $\beta\sim1$, $R_{\rm M}=300$~km, and $\dot{m}_0=20$ (50), we obtain $R_{\rm ph}\simeq 2000$~km ($2\times10^4$~km).  Similar values of the photospheric radius can be obtained by requiring $\tau_{//}=1$, if $\theta_{\rm c}\sim \pi/4$. We note that other opening angles lead to non-spherical photospheres by introducing a dependence of $R_{\rm ph}$ on $\theta_{\rm c}$. 
Any information about changes in the inflow due to the Lense-Thirring torque must be communicated to all parts of the outflow within the photosphere, for it to precess out to that radius. It is possible, however, that the outflow becomes supersonic at $R\ll R_{\rm ph}$, thus becoming decoupled from the inflow. A lower limit on the predicted precession period can be derived, if the outflow remains subsonic within the spherization radius. In this case, $R_{\rm out}\gtrsim R_{\rm sph}$. An accurate calculation of $P_{\rm prec}$ requires detailed knowledge of the outflow thermodynamical properties  and geometry (i.e., density and pressure radial profile, opening angle) and, as such, lies beyond the scope of this paper \citep[see also][]{2019arXiv190502731M}.

Assuming that the observed X-ray flux decay is associated with Lense-Thirring precession, the corresponding period should be longer than a year, since we have only observed one such event within one year of continuous monitoring. For the current spin period $P_{\rm NS}=16$~s, $R_{\rm M}=300$ km, and $\dot{m}_0=20$ (50) we find\footnote{This calculation is just introduced as an order of magnitude estimation. For a discussion about the uncertainties,
we point the reader to \citet{2018MNRAS.475..154M,2019arXiv190502731M}.} $P_{\rm prec}\simeq2.3$~yr (300 yr), if $R_{\rm out}=R_{\rm ph}$ and $P_{\rm prec}\simeq0.4$~yr (3.0 yr), if $R_{\rm out}=2R_{\rm sph}$.
As the spin-up of the NS continues, shorter $P_{\rm prec}$ are expected. Due to the strong dependence of $P_{\rm prec}$ on $\dot{m}_0$, we can place upper limits on the latter by detecting future periodic dips in the X-ray flux due to precession. Future monitoring observations could provide crucial insights on this phenomenon.

\section{Conclusions}
We have analysed monitoring observations of \ngc\ obtained with \swift/XRT within 2018, and presented an updated X-ray light-curve. 
We showed that within a 100 days period the observed flux of the system rapidly decayed. 
Moreover, we triggered \nicer\ target of opportunity observations to follow the spin evolution on the NS during the low-flux epoch.
We showed that the NS of \ngc\ continues to spin up with a rate that translates to a constant mass accretion rate within 2018, even at epochs where the observed flux dropped by a factor of $\sim50$. 
We interpreted the changes in the observed flux as a result of increased absorption and obscuration. Outflows from a radiation-dominated accretion disc  can provide an optically thick structure that could be responsible for the increased absorption.
In this regime, the observed $L_X$ under-predicts the mass accretion rate assuming typical radiative efficiency for the accretion column, thus no strong beaming (if any) is needed to explain the observed super-Eddington luminosity. Based on the inferred properties of \ngc\, we expect the orientation of the outflows to change on year-long timescales due to Lense-Thirring precession. The detection of multiple (quasi-periodic) dips in the X-ray flux within the next decade will provide a firm confirmation for the Lense-Thirring precession being the mechanism responsible for the X-ray obscuration. 

\section*{Acknowledgements}
The authors would like to thank the anonymous referee for their constructive review.
GV would like to thank D. Walton, F. F\"urst, M. Bachetti, and M. Heida for discussions on \ngc\ that took place at the HEAD17 meeting.  
MP acknowledges support from the Lyman Jr.~Spitzer Postdoctoral Fellowship and Fermi Guest Investigator grant 80NSSC18K1745.
We thank Z.~Arzoumanian and the NICER team for their help and assistance during the execution of the NICER ToO observations. 
NICER work at NRL is funded by NASA.
We acknowledge the use of public data from the \swift\ data archive.




\bibliographystyle{mnras}
\bibliography{general}







\bsp	
\label{lastpage}
\end{document}